\documentclass[a4paper]{jpconf}
\usepackage{graphicx}
\begin{document}
\title{The feeding of activity in galaxies: a molecular line perspective}

\author{Santiago Garc\'{\i}a-Burillo$^1$, Francoise Combes$^2$}

\address{$^1$, OAN, Observatorio de Madrid, Alfonso XII, 3, 28014, Madrid, Spain}
\address{$^2$, OAN, Observatoire de Paris, LERMA, CNRS, 61, Av de l«Observatoire, 75014, Paris, France}

\ead{s.gburillo@oan.es, francoise.combes@obspm.fr}

\begin{abstract}

What are the main drivers of activity in the local universe? Observations have been instrumental in identifying the mechanisms responsible for fueling activity in galaxy nuclei. In this context we summarize the main results of the NUclei of GAlaxies (NUGA) survey. The aim of NUGA is to map, at high resolution and high sensitivity, the distribution and dynamics of the molecular gas in the central kiloparsec region of 25 galaxies, and to study the different mechanisms responsible for gas fueling of low-luminosity AGNs (LLAGN). Gas flows in NUGA maps reveal a wide range of  instabilities. The derived gravity torque maps show that only $\sim$1/3 of NUGA galaxies show evidence of ongoing fueling. Secular evolution and dynamical decoupling are seen to be key ingredients to understand the AGN fueling cycle. We discuss the future prospects for this research field with the advent of  instruments like the Atacama Large Millimeter Array (ALMA).

\end{abstract}

\section{The fueling problem}

Active Galactic Nuclei (AGN) must be  fueled with material which lies originally far away from the gravitational influence of the supermassive black hole. On its way to the sphere of influence  of the central engine the gas must lose virtually all of its angular momentum during the process.
The feeding requirements, $dM/dt$ in M$_\odot$ yr$^{-1}$ units, are markedly different in high-luminosity AGNs (QSOs: 10--100M$_\odot$ yr$^{-1}$) and low-luminosity AGNs (LLAGNS: Seyferts and LINERs: 10$^{-5}$--10$^{-2}$M$_\odot$ yr$^{-1}$) [1]. However, observational studies have shown that while in high-luminosity AGNs, the most demanding in terms of feeding requirements,  kiloparsec-scale perturbations produced by bars and galaxy interactions are clearly related to the onset of activity [2], such correlation is marginal if any in the case of  LLAGNs [3] [4] [5]. The search of a ÔuniversalÕ feeding agent in LLAGNs is seen to be challenging. Large-scale bars, bars within bars, lopsided $m=1$ instabilities, warps, nuclear spirals, and  winds from stars, among other mechanisms,  have been invoked as possible feeding agents in LLAGNs. On the other hand, there is growing evidence, based on theoretical models and state-of-the-art numerical simulations that several mechanisms, and not a single mechanism, might be at work and that these operate at different spatial scales [6] [7] [8].   To further complicate this picture in the case of LLAGNs, the duty cycle of activity is expected to be very short and the associated feeding event could be of chaotic/intermittent nature [9][6]. While various modelsÕ predictions are still debated in the literature, there is still ample room for improvement in the picture drawn from observations.

\begin{figure*}[tb!]
   \centering
   \includegraphics[width=7.75cm, angle=0]{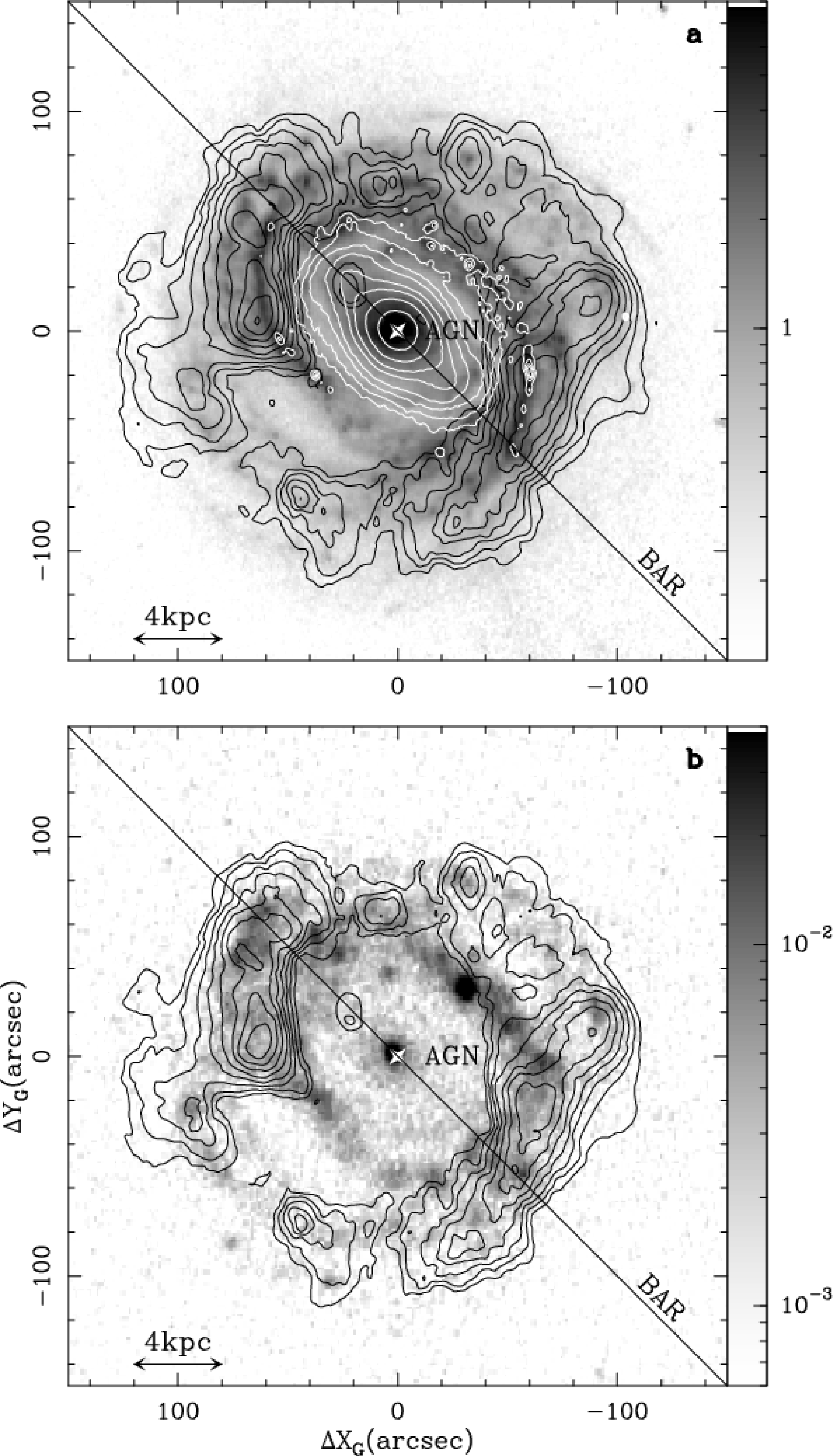}
   \includegraphics[width=8cm, angle=0]{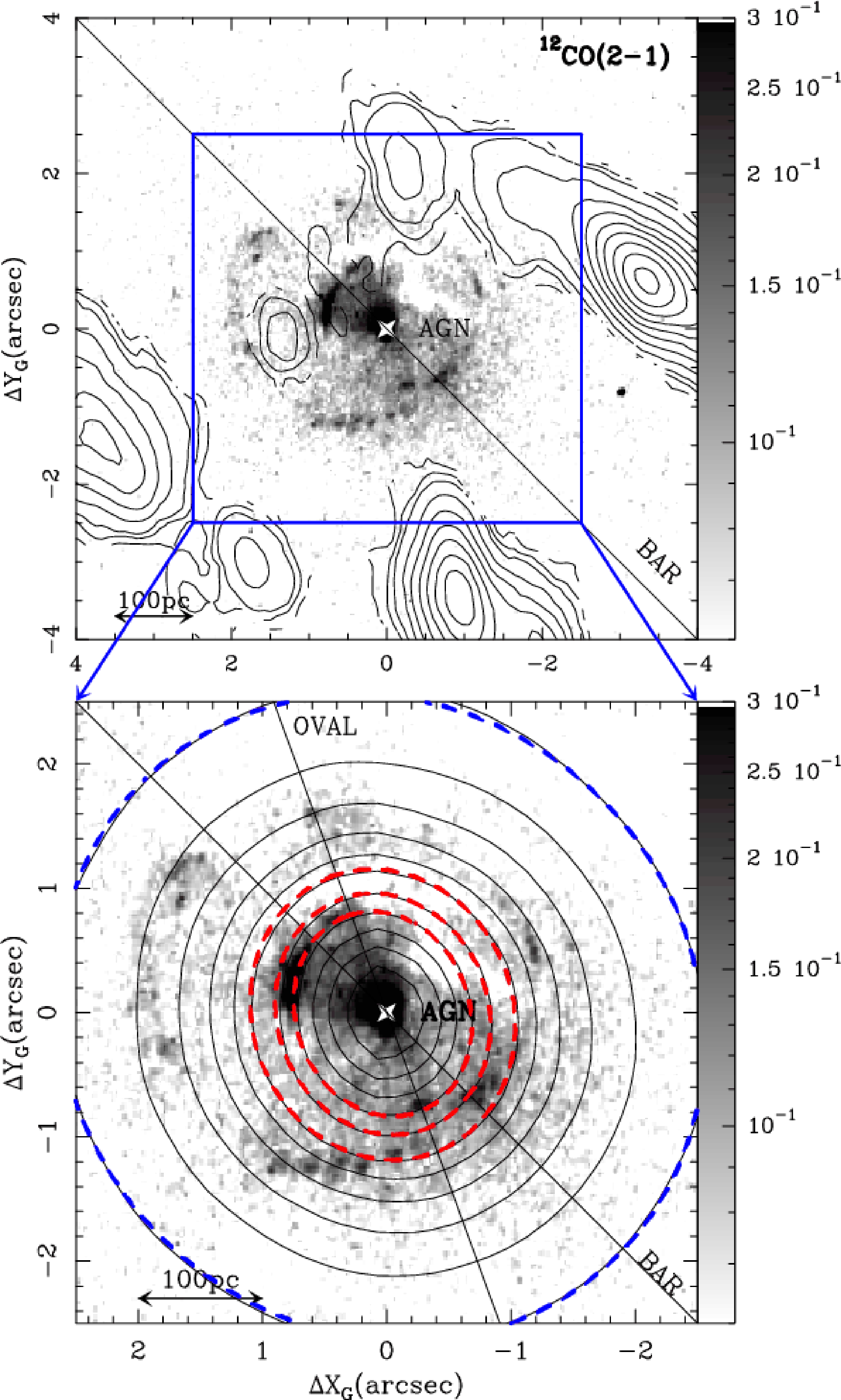}   
   \caption{ a)~({\it Left upper panel}) Overlay of the HI map (black contours), the 3.6 m Spitzer/IRAC image (white contours) and the Spitzer/IRAC image (grey scale) of NGC~4579. The large-scale bar detected in IRAC image-BAR- is identified. A pseudo-ring is detected at UHR in HI emission. b)~({\it Left  lower panel}) Same as a) but HI intensity map (in contours) superposed on the FUV  image (grey scale).  c)~({\it Right upper panel}) The $^{12}$CO(2--1) NUGA map of NGC~4579 is overlaid on the UV image of the central region (in grey scale). d)~({\it Right lower panel}) The K-band contours are overlaid on the UV-image of the galaxy. We show the orientation of the bar and the oval. All images have been deprojected onto the galaxy plane.  Figure adapted from [14].} 
\end{figure*}

\section{An observational perspective: The NUclei of GAlaxies (NUGA) survey}

The NUclei of GAlaxies (NUGA) survey is a high-spatial resolution ($\sim$0.5"-1Õ") and high-dynamic range interferometer CO 
survey of 25 nearby LLAGNs conducted with the IRAM array [10][11]. The aim of this survey is to probe the critical scales for angular momentum transfer  ($<$10--100~pc at $D$$\sim$5--30~Mpc) in a significant sample of galaxies that include Seyferts, LINERs and transition objects. The high sensitivity/resolution of these observations make possible a detailed study of the distribution and kinematics of molecular gas in the circumnuclear disks of these galaxies.  Molecular line maps are used to track down evidence of ÔongoingÕ feeding.

Together with the CO NUGA maps, we also have access to high-resolution NIR maps obtained by HST, Spitzer and/or ground-based telescopes for these galaxies. The NIR imaging is used to derive the stellar potentials, which are combined with the CO maps, to derive the gravity torque budget in the circumnuclear disks of the NUGA targets.  We implicitly assumed that NIR maps are good estimates of mass distribution and that the overall torque budget on the gas is mostly determined by the stellar potential. Two-dimensional torque maps are used  to estimate  the torques averaged over the azimuth $t(r)$ using the gas column density $N(x,y)$ derived from CO as weighting function [12]. The final output of this analysis is the radial profile of the angular momentum transfer efficiency in the disks ($\Delta L$/$L$). With this information at hand we can quantitatively evaluate if there is ongoing AGN fueling in the disks of the galaxies analyzed, down to the spatial resolution of our observations.

\section{NUGA results: statistics}

About one third of the LLAGNs analyzed in NUGA show negative torques $t(r)$$<$0, indicative of inflow, down to typical radial distances $r$$<$25--100~pc. Among these galaxies, which can be considered as those showing 'smoking gun'  evidence of ongoing fueling,  we distinguish two types of  objects:

\begin{figure}[tb!]
   \centering
   \includegraphics[width=16cm, angle=0]{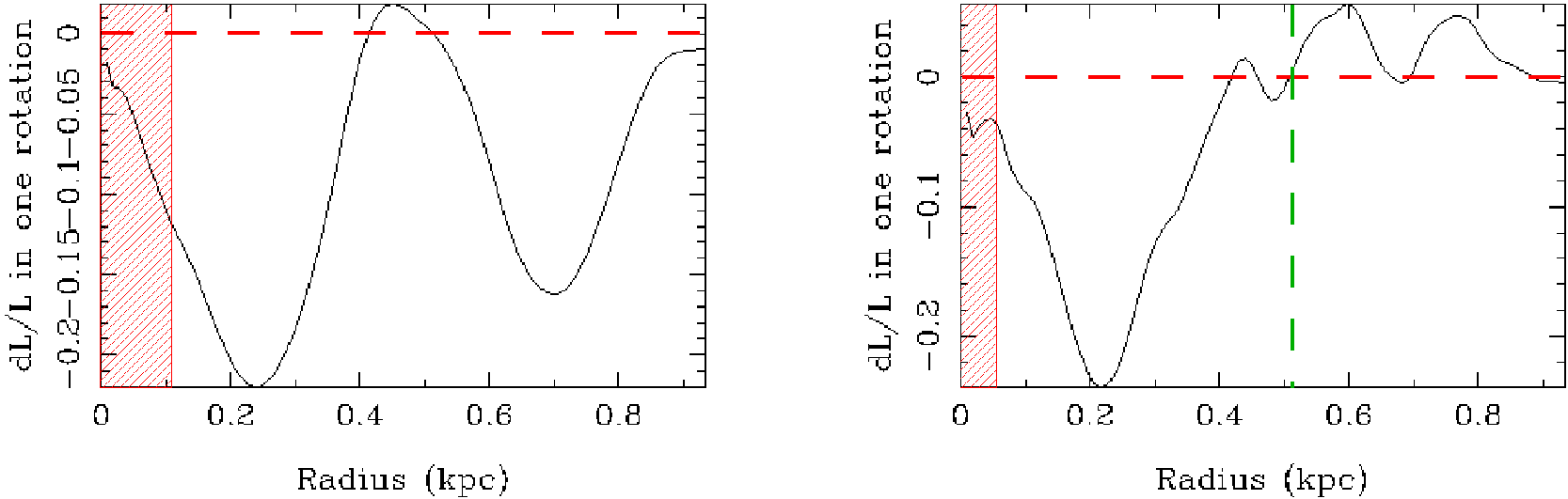}
   \caption{ The radial profile of the angular momentum transfer $\Delta L$/$L$ in the disk of NGC~3627, computed with the HST-NICMOS F160W image is plotted for $^{12}$CO(1--0) (left) and $^{12}$CO(2--1) (right). The (red) dashed region identifies the radial range imposed by the resolution limit  of observations. Torques are systematically negative inside $r$$\sim$400~pc. Figure adapted from [15]. } 
\end{figure}

\begin{itemize}

\item
{\it  \underline{Galaxies showing dynamical decoupling of embedded structures}}:

These include nuclear bars - within - bars/ovals (e.g.; NGC~2782) and nuclear ovals - within -bars  (e.g.; NGC~4579).
The stellar potential of NGC~2782, analyzed by [13]  shows two embedded bars: an outer (weak) oval of $\sim$6~kpc diameter and a (strong)  nuclear bar of $\sim$1.5~kpc diameter. The nuclear bar shows signs of decoupling. This configuration of the stellar potential has facilitated the inflow of molecular gas inside the Inner Lindblad Resonance  (ILR) of the oval. The derived gravity torques $t(r)$  are systematically negative down to $\sim$100--200~pc. The stellar potential  of NGC~4579, studied by [14], shows two embedded bars: an outer bar of $\sim$12~kpc diameter and an embedded weak nuclear oval of $\sim$0.3~kpc diameter. In the outer disk, the decoupling of the spiral arms allows the gas to efficiently populate the Ultra Harmonic Resonance (UHR) region of the large-scale bar. This favors net gas inflow on intermediate scales. Furthermore, closer to the AGN, gas feels negative torques due to the combined action of the outer bar and the nuclear oval. The combination of the two $m=2$ modes produces net gas inflow down to $r$$\sim$50~pc, providing inward gas transport on short dynamical timescales ($\sim$1--3 rotation periods) (Fig.~1).

\item
{\it \underline{Galaxies with ILR-free large-scale bars}}:

 These galaxies are characterized by the apparent absence of gravity torque barriers in their nuclei (e.g.; NGC~3627).
The stellar potential of NGC~3627, analyzed by [15] shows one large-scale bar of $\sim$6~kpc diameter. The  bar has no ILR barrier. Molecular gas is concentrated along the leading edges of the bar and shows no ring feature. Down to the spatial resolution of these observations ($\sim$25~pc), gravity torques $t(r)$ are systematically negative in the circumnuclear disk of  NGC~3627 (Fig.~2). This scenario suggests that the bar in this galaxy is  young and rapidly rotating, and thus has not yet formed an ILR. 

\end{itemize}

On the contrary, about two thirds of the LLAGNs analyzed in NUGA show positive torques $t(r)$$>$0, indicative of outflow,  down to typical radial distances $r$$<$300~pc. This {\it puzzling} gravity torque budget is found in two categories of objects in our sample:

\begin{itemize}

\item
{\it \underline{Galaxies showing Ônon-cooperativeÕ embedded structures}}:

 In these galaxies, nuclear bars/ovals - within - bars, like those found in NGC~4321 and NGC~6951 are not
always conducive to gas inflow at present [12][16][17]. The stellar potential of NGC~6951, analyzed by [12] and [16]
 shows two embedded bars: a large-scale bar of $\sim$8~kpc diameter and an inner oval of $\sim$0.4~kpc diameter. Molecular gas is stalled in the ILR ring of the bar.  While molecular gas is also detected at the AGN locus [12][17][18], gas is gaining angular momentum inside the ILR due to oval forcing (Fig.~3).

\item
{\it \underline{Galaxies showing featureless/mostly axisymmetric potentials}}: 

The absence of a clear non-axisymmetric feature in  the stellar potential makes the angular
momentum transfer in the nucleus a very inefficient process in some galaxies at present (e.g.; NGC~4826 [11]; NGC~7217 [16][19]; NGC~5953 [20]).

\end{itemize}

\begin{figure}[tb!]
   \centering
   \includegraphics[width=8cm, angle=0]{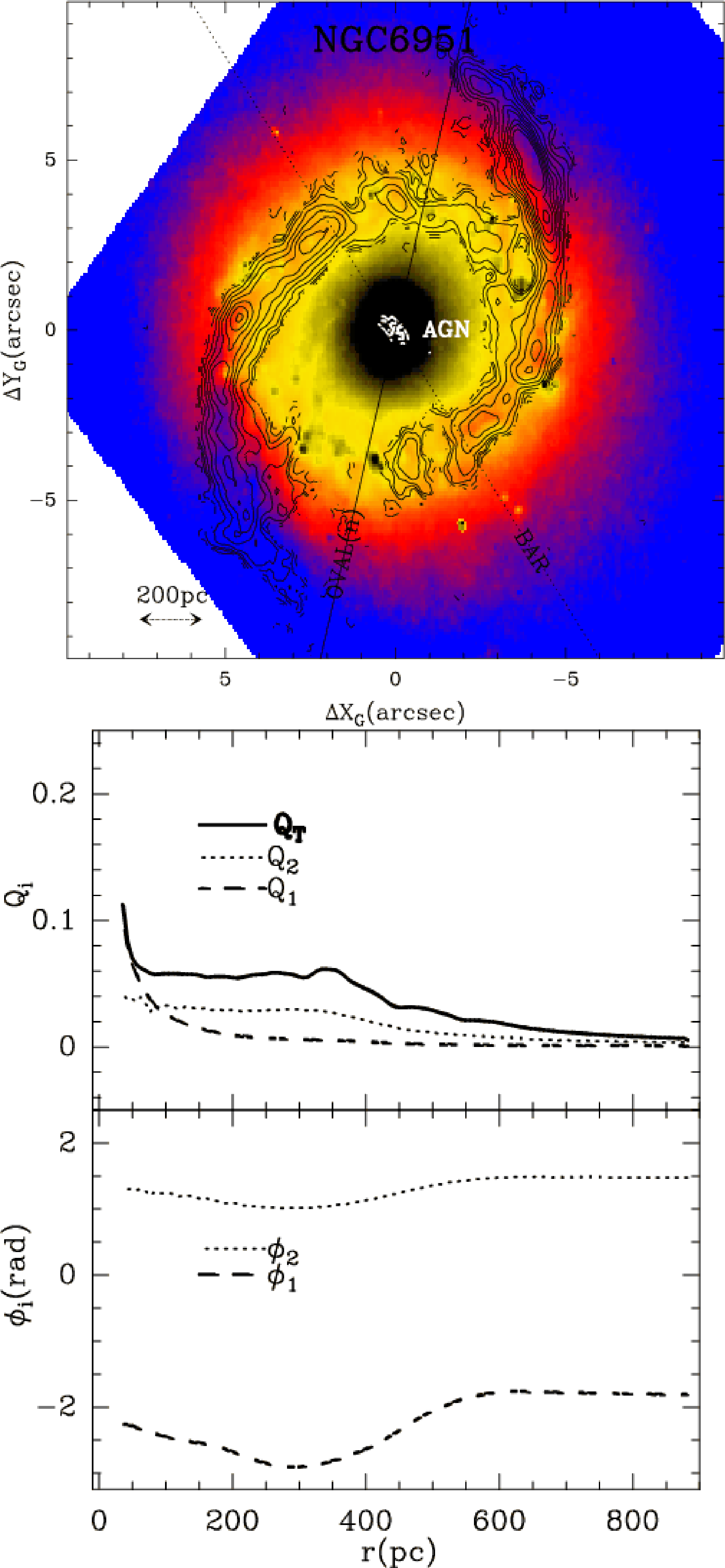}
   \caption{a) ({\it Upper panel}) Overlay of the $^{12}$CO(2--1) NUGA map (contours) on the J-band HST image (grey scale) of NGC~6951;  images have been deprojected onto the galaxy plane. The orientations of the large-scale bar-BAR-and of the nuclear oval-OVAL(n)-are shown. b) ({\it Lower panels}) We plot the strengths ($Q_i$, $i$=1,2) and phases ($\Phi_i$, $i$=1,2) of the $m=1$ and $m=2$ perturbations inside the image field-of-view. Figure adapted from [12]. } 
\end{figure}

In summary, the overall results of NUGA indicate that high spatial resolution is instrumental in quantifying angular momentum transfer processes at critical scales ($\sim$10--100~pc). Gas flows in NUGA maps reveal a wide range of large-scale and embedded $m=2$, $m=1$ instabilities in the circumnuclear disks of LLAGNs. The derived gravity torque maps indicate that gas is frequently stalled in rings, which are the signposts of gravity torque barriers. Only $\sim$1/3 of NUGA galaxies show negative torques down to $\sim$50~pc. Several short-lived ($<$a few 10$^7$~yr) mechanisms are at work to drain angular momentum. These various fueling mechanisms are related to bar cycles [21][22]. Secular evolution and dynamical decoupling are key ingredients to understand the AGN fueling cycle.

\section{Future surveys with ALMA}

To confirm the different feeding scenarios we must improve the spatial resolution and sensitivity of millimeter line observations. 
The advent of the Atacama Large Millimeter Array (ALMA) will allow us to boost the spatial resolution of observations of galaxy nuclei.
The full ALMA capabilities will make possible to map the emission of a set of molecular gas tracers in the inner $\sim$1--10~pc of a significant sample of $\sim$50 nearby ($D<$5--30~Mpc) AGNs in about 100~hrs of observing time with an order of magnitude higher sensitivity compared to the current NUGA survey.  Improving the statistics will allow us to better explore the evolutionary sequence in AGN fueling.

The CO and NIR maps of the observed galaxies, obtained at a common spatial resolution of $\sim$0.05"--0.1" with ALMA and HST/VLT, will provide a sharp view of the gravity torque budget in the circumnuclear disks. The efficiency of new feeding mechanisms will be observationally tested at unprecedented scales $\sim$1--10~pc (see F. Combes' contribution to this symposium). In particular, the role of $m=1$ modes, gas self-gravity, granularity of the stellar potential, and dynamical friction of GMCs in AGN fueling will be thus tested. 

Another key ingredient in galaxy evolution is AGN feedback. ALMA enhanced capabilities will also allow us to simultaneously search for the signature of molecular gas outflows and high-velocity winds in nearby AGNs.

\smallskip

\end{document}